%% file: main.tex
\documentclass[conference]{IEEEtran}
\IEEEoverridecommandlockouts
\usepackage{lipsum}
\usepackage{cite}
\usepackage{amsmath,amssymb,amsfonts}
\usepackage{balance}
\usepackage{algorithm}
\usepackage{subcaption}
\usepackage{relsize}
\usepackage{braket} 
\usepackage{algpseudocode} 

\usepackage{graphicx}
\usepackage{textcomp}
\usepackage{xcolor}
\def\BibTeX{{\rm B\kern-.05em{\sc i\kern-.025em b}\kern-.08em
    T\kern-.1667em\lower.7ex\hbox{E}\kern-.125emX}}

\renewcommand{\baselinestretch}{0.95}
\begin{document}

\title{BiBiEQ: \underline{Bi}variate \underline{Bi}cycle Codes on \underline{E}rasure \underline{Q}ubits\\
}

\author{
\IEEEauthorblockN{Ameya S. Bhave\IEEEauthorrefmark{1}, Navnil Choudhury\IEEEauthorrefmark{2}, Andrew Nemec\IEEEauthorrefmark{1}, Kanad Basu\IEEEauthorrefmark{2}}
\IEEEauthorblockA{\IEEEauthorrefmark{1}Department of Electrical and Computer Engineering, The University of Texas at Dallas, Richardson, TX, USA}
\IEEEauthorblockA{\IEEEauthorrefmark{2}Department of Electrical, Computer, and Systems Engineering, Rensselaer Polytechnic Institute, Troy, NY, USA}
}

\maketitle

\input{sections/abstract/Abstract}
\input{sections/introduction/Introduction}
\input{sections/background/Background}

\input{sections/methodology/Methodology}

\input{sections/performance/Performance}
\input{sections/conclusion/Conclusion}

\end{document}

%% file: sections/abstract/Abstract.tex
\begin{abstract}
Erasure qubits reduce overhead in fault-tolerant quantum error correction (QEC) by converting dominant faults into detectable errors known as erasures. They have demonstrated notable improvements in thresholds and scaling in surface and Floquet code memories. In this work, we use erasure qubits on Bivariate Bicycle (BB) codes from the quantum low-density parity-check (QLDPC) regime. Owing to their sparse structure and favorable rate–distance trade-offs, BB codes are practical candidates for QEC. We introduce BiBiEQ, a novel framework that compiles a given BB code into an erasure-aware memory circuit $C_E$. This erasure circuit $C_E$ comprises erasure checks (ECs), resets, and erasures spread over a user-specified erasure check schedule (2EC, 4EC). BiBiEQ converts this erasure circuit $C_E$ into the stabilizer circuit $C$ for general-purpose decoding. BiBiEQ provides two engines for this conversion, BiBiEQ-Exact and BiBiEQ-Approx. BiBiEQ-Exact preserves the joint-erasure correlations and serves as our accuracy benchmark, while BiBiEQ-Approx uses an independence approximation to accelerate large sweeps and expose accuracy–throughput trade-offs. Using BiBiEQ, we decode the stabilizer circuits to get a per-round logical error rate (LER) for the BB codes and quantify the effect of the EC schedules on the correctable operating region below the pseudo-threshold. The 4EC schedule keeps the accuracy of both engines close to one another, making BiBiEQ-Approx a reliable proxy for BiBiEQ-Exact for faster sweeps. Below the pseudo-threshold, the code distance ($d$) hop from $d{:}6\!\to\!10$ yields a drop in LER by $\approx10$-$17\times$ larger than \(d{:}10\!\to\!12\), showing that most gains are realized by \(d=10\).
\end{abstract}


%% file: sections/introduction/Introduction.tex
\section{Introduction}
\label{intro}
Quantum error correction (QEC) is essential for scalable quantum computing. The dominant cost of scalability is the overhead from extra qubits, checks, and cycles required for QEC. In 2D-local architectures, this cost grows quadratically with code distance and becomes the main bottleneck. Erasure qubits provide a promising route to low-overhead quantum error correction by converting certain hardware faults into detectable erasures, thereby revealing error locations \cite{b3}. In the case of dual-rail hardware, a photon loss or leakage in either rail produces an invalid occupancy pattern that a photon-number or parity check flags as an erasure \cite{b1}. In contrast, the standard circuit-level model assumes stochastic Pauli faults with unknown location, increasing the inference burden on the decoder \cite{b8}. An erasure-biased approach to QEC has been shown \cite{b2, b3}  to raise thresholds and improve the subthreshold scaling of logical error rates, particularly in fault-tolerant quantum memory experiments with surface and Floquet codes. However, these advantages come at a cost of additional circuit operations like erasure checks (ECs) and resets \cite{b2, b3}. Each erasure check and reset provides information by flagging fault locations, but also introduces additional error channels. Increasing the EC frequency increases both the available information and the induced noise. Thus, the key design lies in balancing the information gained from erasure detection against the circuit overhead it incurs. 

Surface codes have been the main testbed for exploring erasure-aware implementations to understand the interplay with circuit-level constructions such as entangling schedules and check weights \cite{b1,b2}. However, surface codes have a poor encoding rate, and increasing the distance makes the overhead grow quadratically \cite{b9}. Extra ECs and resets deepen cycles and raise measurement load. Subsystem variants reduce check weight using gauges but need multiple gauge-measurement rounds per stabilizer, which increases cycle count and circuit depth when ECs are performed \cite{b7}. These limitations motivate exploring erasure-aware implementations of more scalable architectures, such as quantum low-density parity-check (QLDPC) codes, which currently remain underexplored.

\begin{figure}[!t]
  \centering
  \includegraphics[width=\linewidth, keepaspectratio]{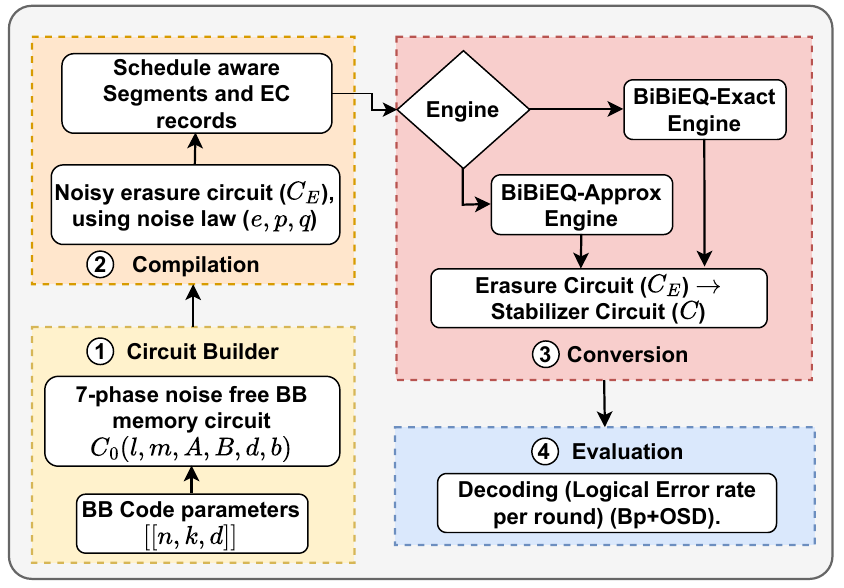}
  \caption{BiBiEQ framework overview: (1) Circuit builder from code parameters [[$n,k,d$]]. (2) Compilation to apply the noise law and collect schedule-aware segments. (3) Convert $C_{E}$ into a stabilizer circuit $C$. (4) Evaluation to get LER.}
  \label{fig:bibieq_framework_overview}
  \vspace{-3mm}
\end{figure}

QLDPC codes use sparse, bounded-degree checks and can achieve nonzero rate with growing distance, thereby lowering the overhead per logical qubit \cite{b6,b10}. This sparsity pairs well with erasure qubits because with the erasure set \(E\) known, decoding reduces to solving local parity constraints on \(E\), instead of searching over the full code. Among QLDPC families, Bivariate Bicycle (BB) codes offer surface-code-like logical performance at a higher encoding rate and admit regular low-depth entangling schedules, which makes EC interleaving cheaper \cite{b4}. These properties make BB codes a natural target for erasure-aware circuits.

In this paper, we introduce BiBiEQ, a novel framework to implement erasure qubits over BB codes. Figure~\ref{fig:bibieq_framework_overview} summarizes the BiBiEQ pipeline which comprises \textbf{four} main stages: 

\noindent\textbf{1. Circuit Builder}
A memory circuit is instantiated in \texttt{Stim} from the BB code parameters~\cite{b11} (see  Section~\ref{methd:subsec:code_spec_circ_build}).

\noindent\textbf{2. Compilation}
The noise law is applied to obtain an erasure-annotated circuit. Each wire is partitioned into \emph{segments} by resets (see Section~\ref{subsec:noise_erasure_compile}).

\noindent\textbf{3. Conversion}
Two conversion engines are used—BiBiEQ-Exact and BiBiEQ-Approx—to map the erasure-annotated circuit to a stabilizer circuit suitable for decoding. Their design and trade-offs are detailed in Section~\ref{subsec:methodology_erasure_approx_exact}.

\noindent\textbf{4. Evaluation}
A belief propagation-based decoder is used to estimate the per-round logical error rate \(p_{L}(e)\), which is then used for performance analysis (see Section~\ref{sec:performance}).

The key contributions of the paper are as follows:
\begin{itemize}
  \item \textbf{BiBiEQ framework with dual engines :}
  We design \emph{BiBiEQ}, an end-to-end, erasure-aware framework for BB codes that segments circuits by resets/EC records, computes segment-wise posteriors, and converts the erasure-annotated circuit to a stabilizer circuit via \textbf{\emph{BiBiEQ-Exact}} (posterior-faithful) and \textbf{\emph{BiBiEQ-Approx}} (tractable approximation), enabling general-purpose decoding.

  \item \textbf{Seven-phase circuits with schedule-aware ECs :}
  We develop a parameterized generator for seven-phase BB memory circuits in \texttt{Stim}. The generator supports schedules, segments, and EC records, turning erasure-aware experiments into a reproducible benchmark.

  \item \textbf{2EC vs.\ 4EC schedule performance :}
    We evaluate the performance of 2EC versus 4EC schedules under an erasure-biased noise law \((e,p,q)\) with $p$ = $0.1e$ and $q$ = $e$, and we quantify the information–overhead trade-off using the logical error rate (LER) per round (syndrome measurement cycle).

  \item \textbf{Subthreshold-scaling analysis for the BB family :}
  We characterize subthreshold scaling across distances using the subthreshold scaling law to show how an increase in distance translates into practical LER gains.
\end{itemize}

%% file: sections/background/Background.tex
\section{Background}
\vspace{-1ex}
\input{sections/background/erasure_qubits/erasure_qubits}
\vspace{-1ex}
\input{sections/background/bivariate_bicycle_code/bivariate_bicycle_code}
\vspace{-1ex}
\input{sections/background/segment_abstraction/segment_abstraction}

%% file: sections/background/erasure_qubits/erasure_qubits.tex
\subsection{Erasure Qubits}\label{subsec:bg_erasure_qubits}

\begin{figure}[!t]
  \centering
  \includegraphics[width=.9\linewidth, height=7.5cm]{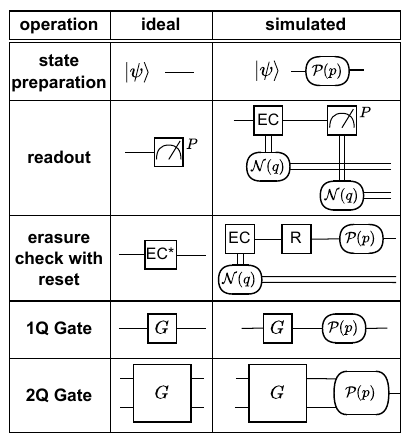}
  \caption{Ideal vs.\ simulated operations under the erasure model. Each simulated block appends an erasure/Pauli channel with probabilities $p$ or $q$, and EC triggers a reset $R$.}
  \label{fig:erasure_protocol}
  \vspace{-4mm}
\end{figure}

Erasure qubits expose the \emph{location} of faults with high reliability. When a qubit leaves the computational subspace, a herald is produced, and the spacetime location is known. Decoders then treat flagged events as erasures rather than unknown Pauli faults, simplifying inference and improving thresholds/sub-threshold scaling in topological settings. Such settings also consider \emph{imperfect} heralding—false positives/negatives on the erasure flag—since practical probes and readout are noisy \cite{b1,b2,b3}. Figure~\ref{fig:erasure_protocol} represents a simulation convention introduced in \cite{b2, b3} that converts an ideal circuit into an erasure-annotated circuit for simulation. Each ideal operation is accompanied by a local fault generator and an erasure check, as explained below:
\begin{itemize}
  \item \textbf{State preparation :} An ideal $\ket{\psi}$ is followed by a depolarizing channel, $\mathcal{P}(p)$, with error rate $p$.
  \item \textbf{Readout :} Immediately before a measurement, an erasure check is inserted that probes whether the data qubit is erased/leaked. The erasure flag itself is subject to classical bit-flip noise $\mathcal{N}(q)$ with rate $q$ to capture imperfect heralding. The subsequent measurement $M^{P}$ in Pauli basis $P\!\in\!\{X,Z\}$. A raised flag marks the qubit as \emph{erased} at that time step.
  \item \textbf{Erasure check with reset :} Schedules perform an explicit mid-circuit EC. If a flag is raised, a reset $R$ is applied to return to the computational subspace. A post-reset depolarizing channel $\mathcal{P}(p)$ models re-initialization error.
  \item \textbf{1Q/2Q gates :} After each gate $G$, a depolarizing channel $\mathcal{P}(p)$ is applied to the associated qubits. Entangling gates can spread Pauli faults, but flagged erasures retain their known locations and are exploited at decoding time \cite{b2}.
\end{itemize}

%% file: sections/background/bivariate_bicycle_code/bivariate_bicycle_code.tex
\subsection{Bivariate Bicycle (BB) Codes within QLDPC Family}
\label{bg:subsec:bg_bivariate_bicycle_code}

Quantum LDPC (QLDPC) codes exploit sparse Tanner graphs to enable scalable syndrome extraction. Among these, the
\emph{Bivariate Bicycle (BB)} construction uses a pair of bivariate polynomials $(A(x,y),B(x,y))$ over $\mathbb{F}_2$ to realize a toroidal Tanner graph with two interleaved data-qubit lattices and weight-six $X/Z$ checks \cite{b4}. BB codes preserve LDPC sparsity while offering strong finite-length performance under BP+OSD decoding and hardware-friendly layouts \cite{b5}.

\begin{figure}[!t]
  \centering
  \includegraphics[width=.8\linewidth,keepaspectratio]{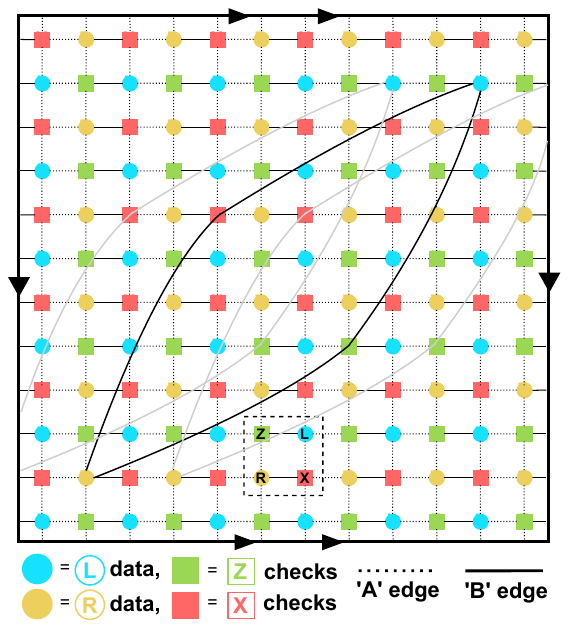}
  \caption{Tanner graph of the $[[72,12,6]]$ BB code. Two interleaved data lattices (blue/yellow) connect to sparse $Z$ (green) and $X$ (red) checks. Arrows show toroidal wraparound. Monomials in $A$ generate dotted ``A'' edges, and monomials in $B$ generate solid ``B'' edges.}
  \label{fig:bb_tanner_72_12_6}
  \vspace{-4mm}
\end{figure}

Figure~\ref{fig:bb_tanner_72_12_6} shows the \emph{Tanner graph} of a BB code $[[72,12,6]]$, on an $(l,m)=(6,6)$ torus. With two $l\times m$ data lattices, the block length becomes $n = 2lm$ and we get $n=72$. We place the code on a torus $\mathbb{Z}_l \times \mathbb{Z}_m$ so indices wrap around in both coordinates. Blue/yellow circles denote the data qubits on the two interleaved lattices \(L\) and \(R\) while green/red squares represent \(Z\)/\(X\) checks. Dotted “A” edges come from monomials of \(A(x,y)\) and solid “B” edges come from \(B(x,y)\).  A binary bivariate polynomial
$C(x,y)=\sum_{(a,b)\in S_C} x^{a}y^{b}$ defines a \emph{displacement set} $S_C \subseteq \{0,\ldots,l\!-\!1\}\times\{0,\ldots,m\!-\!1\}$. We index unit cells by \(i=(u,v)\in\Lambda\), where $\Lambda$ is the index set 
with \(0\le u<l\), \(0\le v<m\). Each cell \(i\) hosts the two data qubits \(q(L,i), q(R,i)\) and two ancilla \(q(X,i), q(Z,i)\) qubits on the two sublattices as shown by the dashed square box in Figure~\ref{fig:bb_tanner_72_12_6}. For any displacement $\delta=(\Delta x,\Delta y)$ we use modular addition using Eq.~(\ref{eq:modular_lattice_addition}):
\begin{equation}
i \oplus \delta := \big((u+\Delta x)\bmod l,\ (v+\Delta y)\bmod m\big),
\label{eq:modular_lattice_addition}
\vspace{-1.5mm}
\end{equation}
Given $S_A$= \{(3,0),(0,1),(0,2)\} and $S_B$=\{(0,3),(1,0),(2,0)\}, enumerating $s^A_k\!\in\!S_A$ and $s^B_k\!\in\!S_B$ for $k=1,2,3$ it defines
$A_k(i)=i\oplus s^A_k, \ B_k(i)=i\oplus s^B_k$ and the transpose to be $A_k^{\top}(i)=i\oplus(-s^A_k), \ B_k^{\top}(i)=i\oplus(-s^B_k)$. A $Z$-check at cell $i$ connects to $L$-data at the neighbors $A_k(i)$ and to $R$-data at the neighbors $B_k(i)$ (and analogously for $X$-checks with Pauli types swapped). For example with \((l,m)=(6,6)\), \(i=(2,4)\), and \(A_2=(0,1)\), we get \(A_2(i)=(2+0,4+1)\), with \(A_3=(0,2)\), the reverse step is \(A_3^{\top}(i)=(2-0,4-2)\). The per-check weight is therefore $|S_A|+|S_B|=3+3=6$. Fixing $(A,B)$ preserves local geometry and check weight, while varying $(l,m)$ scales $n$ and the distance $d$, giving a controlled distance ladder used later in our BiBiEQ experiments.

%% file: sections/background/segment_abstraction/segment_abstraction.tex
\subsection{Segment abstraction for erasure-aware execution}
\label{subsubsec:bg-segments}

\begin{figure}[!t]
  \centering
\includegraphics[width=.9\linewidth,keepaspectratio]{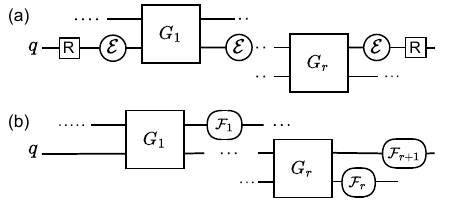}
  \caption{Segment abstraction. (a) A segment $s$ over qubit $q$ in an erasure circuit $C_{E}$. (b) The associated spacetime locations are realized as fault channels $\mathcal{F}_i$ in the stabilizer circuit.}
  \label{fig:segments}
  \vspace{-4mm}
\end{figure}

In the erasure-qubit view, each wire is partitioned by resets into segments.
A segment is the open interval between consecutive resets.
Erasure checks (EC) occur inside a segment.
A segment ends only when a flagged check is followed by a reset (EC$^{*}$).
Thus, the EC-plus-reset policy fixes the segment boundaries.
Figure~\ref{fig:segments} presents the segment abstraction. Figure~\ref{fig:segments}(a) depicts wire $q$ inside the \emph{erasure circuit} $C_{E}$. Circles $\mathcal{E}$ mark erasure locations inserted between consecutive operations. The two-qubit entangling gates within the segment are represented as $G_1,\ldots, G_r$. The segment ends at the next reset. After each operation, the depolarizing channel $\mathcal{P}(p)$ is applied. Each EC or measurement bit then passes through the binary symmetric channel $\mathcal{N}(q)$.

Figure~\ref{fig:segments}(b) shows the conversion to a stabilizer circuit $C$. Inside a segment the unitary is a sequence of Clifford layers (e.g., CNOT gates). By Pauli conjugation, any fault drawn from $\mathcal{P}(p)$ can be pushed forward to a canonical set of post-interaction sites that follow each entangling gate $G_i$ and at the segment end. Hence, it replaces the erasure-annotated circuit $C_{E}$ by a stabilizer circuit conditioned on the EC record within the segment (i.e., which $\mathcal{E}$ sites are flagged). The Pauli faults are injected only at these sites $\{F_1,\ldots, F_{r+1}\}$, yielding an equivalent distribution of syndromes and measurement outcomes for decoding. Resets prevent propagation across segments, and the bit flips are modeled by $\mathcal{N}(q)$ \cite{b1,b2}.


%% file: sections/methodology/Methodology.tex
\section{Methodology}\label{sec:methodology}

\input{sections/methodology/overview_design_goals/overview_design_goals}

\input{sections/methodology/code_specification_circuit_build/code_specification_circuit_build}

\input{sections/methodology/parameterized_noise_erasure_compilation/parameterized_noise_erasure_compilation}

\input{sections/methodology/erasure_approx_exact_injection/erasure_approx_exact_injection}

%% file: sections/methodology/overview_design_goals/overview_design_goals.tex
We present \textbf{BiBiEQ}, a novel framework for implementing Bivariate Bicycle (BB) codes under an erasure-qubit noise model with schedule-aware error correction. BiBiEQ unifies four components: $(1)$ A circuit generator builds a seven-phase circuit for an $[[n,k,d]]$ code and defines the logical Pauli operators for the \(k\) logical qubits as explicit measurement parities that are tracked as observables. $(2)$ A parameterized compiler injects physical noise and erasure processes via segment- and schedule-aware transformations \cite{b12}. $(3)$ Sampling engines realize erasure events using \textbf{Exact} and \textbf{Approximate} formalisms \cite{b2,b3}. $(4)$ A decoder-agnostic pipeline interfaces with multiple decoders. These components are detailed in the subsections that follow.

BiBiEQ achieves three design goals. \emph{First}, a \textbf{schedule-agnostic build stage} synthesizes a circuit once and reuses it across sweeps as described in Section~\ref{methd:subsec:code_spec_circ_build}. \emph{Second}, \textbf{engine-agnostic erasure injection} enables head-to-head comparisons between fidelity-preserving exact realizations and higher-throughput approximate variants as explained in Section~\ref{subsec:methodology_erasure_approx_exact}, Algorithm~\ref{alg:bibieq-unified}. \emph{Third}, \textbf{uniform decoding and aggregation} ensure per-round logical error rates are computed by the same pipeline across configurations explained in Sections~\ref{subsubsec:bench_eval}, \ref{sec:perf:metrics}. Together, these properties make BiBiEQ a unified, reproducible framework for assessing BB-code performance under erasure-qubit noise.

%% file: sections/methodology/code_specification_circuit_build/code_specification_circuit_build.tex
\subsection{Code Specification \& Circuit Build}\label{methd:subsec:code_spec_circ_build}

We instantiate BiBiEQ on a family of Bivariate Bicycle (BB) codes \cite{b5,b6}. Table~\ref{tab1:bb_codes_family} represents the codes from the BB code family and summarizes the characteristic properties of each code. Each $[[n,k,d]]$ instance is specified with a fixed polynomial pair $A(x,y)=x^{3}+y+y^{2}$, $B(x,y)=y^{3}+x+x^{2}$ and lattice periods $(l,m)\in\{(6,6),(9,6),(12,6)\}$. For the toroidal two-sublattice layout from Section~\ref{bg:subsec:bg_bivariate_bicycle_code}, the block length is exactly $n=2lm$, giving $n\in\{72,108,144\}$ and encoding rate to be $r=k/n$ as shown in Table~\ref{tab1:bb_codes_family}. We hold the polynomial pair $A(x,y)$ and $B(x,y)$ fixed. By keeping the pair fixed, we preserve the local geometry and check weight so that the changes in performance are reflected by the lattice size and code distance $d$ rather than local connectivity \cite{b4}.

\begin{table}[!t]
\caption{Bivariate Bicycle codes and their parameters.}
\centering
\renewcommand{\arraystretch}{1.3}
\setlength{\tabcolsep}{3pt}
\begin{tabular}{|c|c|c|c|c|}
\hline
\textbf{$[[n,k,d]]$} & \textbf{Encoding rate $r$} & \textbf{$(l,m)$} & \textbf{$A(x,y)$} & \textbf{$B(x,y)$} \\
\hline
$[[72,12,6]]$   & $1/12$ & $(6,6)$  & $x^3 + y + y^2$ & $y^3 + x + x^2$ \\
\hline
$[[108,8,10]]$  & $1/27$ & $(9,6)$  & $x^3 + y + y^2$ & $y^3 + x + x^2$ \\
\hline
$[[144,12,12]]$ & $1/24$ & $(12,6)$ & $x^3 + y + y^2$ & $y^3 + x + x^2$ \\
\hline
\end{tabular}
\label{tab1:bb_codes_family}
\vspace{-4mm}
\end{table}

For the BB polynomial pair, the displacement sets are $S_A=\{(3,0),(0,1),(0,2)\}$ and $S_B=\{(0,3),(1,0),(2,0)\}$, giving per-check weight $|S_A|+|S_B|=6$ of a data qubit, with the “A/B” edge pattern as shown in Figure~\ref{fig:bb_tanner_72_12_6}.  These displacement sets determined by the monomials of $A(x,y)$ and $B(x,y)$ uniquely specify all data–ancilla couplings in the Tanner graph. A syndrome-measurement cycle is achieved by ordering these couplings into CNOT layers that separate $X$- and $Z$-side interactions, using the 7-phase CNOT pattern, as shown in Table~\ref{tab:seven_phase_schedule} \cite{b4}.
Ancilla qubits are arranged as $X$-ancilla \(q(X)\) and $Z$-ancilla \(q(Z)\). Data qubits are arranged on the two bicycle halves \(q(L),q(R)\). \(\{A_k,B_k\}_{k=1}^3\) are  the displacement triples derived from \(A(x,y)\) and \(B(x,y)\). For a unit-cell index \(i = (u, v)\) on an \(l\times m\) grid with wrap-around edges, \(A_k(i)\) and \(B_k(i)\) denote the neighbor cells reached by the \(k\)-th offsets. The transposes \(A_k^{\top}(i), \ B_k^{\top}(i)\) are the reverse offsets used when control and target roles are swapped as previously explained in Section~\ref{bg:subsec:bg_bivariate_bicycle_code}. 
\begin{table}[!b]
\centering
\caption{Seven-phase BB CNOT schedule.}
\renewcommand{\arraystretch}{1.3}
\resizebox{\columnwidth}{!}{%
\begin{tabular}{|c|c|l|l|}
\hline
\textbf{Rd.} & 
\textbf{\begin{tabular}[c]{@{}c@{}}$X$-checks, \\ $Z$-checks \end{tabular}} &
\textbf{\begin{tabular}[c]{@{}c@{}}$X$-side CNOTs \\ (control $\to$ target)\end{tabular}} &
\textbf{\begin{tabular}[c]{@{}c@{}}$Z$-side CNOTs \\ (control $\to$ target)\end{tabular}} \\ \hline
1 & $Z_{1}$ & -- & $q(R, A_{1}^{\top}(i)) \to q(Z,i)$ \\ \hline
2 & $X_{1}$ + $Z_{2}$ & $q(X,i) \to q(L, A_{2}(i))$ & $q(R, A_{3}^{\top}(i)) \to q(Z,i)$ \\ \hline
3 & $X_{2}$ + $Z_{3}$ & $q(X,i) \to q(R, B_{2}(i))$ & $q(L, B_{1}^{\top}(i)) \to q(Z,i)$ \\ \hline
4 & $X_{3}$ + $Z_{4}$ & $q(X,i) \to q(R, B_{1}(i))$ & $q(L, B_{2}^{\top}(i)) \to q(Z,i)$ \\ \hline
5 & $X_{4}$ + $Z_{5}$ & $q(X,i) \to q(R, B_{3}(i))$ & $q(L, B_{3}^{\top}(i)) \to q(Z,i)$ \\ \hline
6 & $X_{5}$ + $Z_{6}$ & $q(X,i) \to q(L, A_{1}(i))$ & $q(R, A_{2}^{\top}(i)) \to q(Z,i)$ \\ \hline
7 & $X_{6}$ & $q(X,i) \to q(L, A_{3}(i))$ & --- \\ \hline
\end{tabular}%
}
\label{tab:seven_phase_schedule}
\end{table}

Each row of the schedule executes a vertex-disjoint layer in which \(q(X,i)\) controls CNOTs onto the specified neighbors in \(q(L)\) or \(q(R)\), while in the same layer the corresponding data qubits control onto \(q(Z,i)\) via the transposed offsets. The six offsets are interleaved across seven rounds so that, by the end of the cycle, every \(q(X,i)\) has acted once on each of its three data neighbors and every \(q(Z,i)\) has received one control from each of its three neighbors. Rounds 1 and 7 idle the complementary side to preserve vertex-disjointness at fixed local degree \(w=6\). This gives us a noise-free seven-phase circuit \(C_{0}(l, m, A, B, d, b)\), where distance $d$ is also the number of rounds (syndrome-measurement cycles) and $b \in \{X, Z\}$ is the logical basis. This circuit dovetails with the erasure-check (EC) schedules as each phase is depth-1, providing clean EC insertion points. Separating $X$- and $Z$-side interactions within each phase enables precise placement of ECs and resets. Consequently, we attribute observed trade-offs to EC design rather than code construction, which sets up the segment-based noise/schedule insertion.

%% file: sections/methodology/parameterized_noise_erasure_compilation/parameterized_noise_erasure_compilation.tex
\subsection{Parameterized Noise and Erasure Compilation}
\label{subsec:noise_erasure_compile}

\begin{figure}[!t]
  \centering
\includegraphics[width=\linewidth,keepaspectratio]{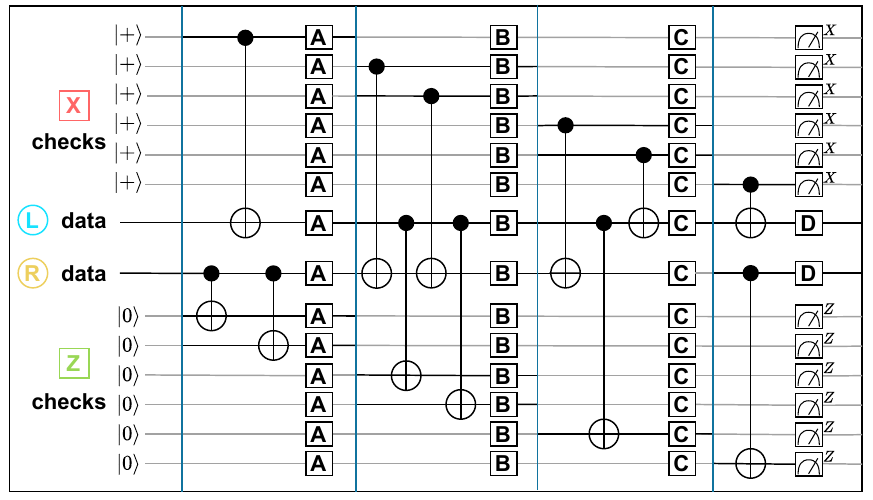}
  \caption{BB memory circuit for syndrome measurement with checkpoints \(A\)–\(D\) (EC\,+\,reset sites).}
  \label{fig:erasure_check_schedule}
  \vspace{-4mm}
\end{figure}
For a particular given schedule-agnostic circuit $C_{0}(l,m,A,B,d,b)$ we add noise using an \emph{erasure-biased} noise law $(e,p,q)$. Here $e$ denotes the per-site erasure probability sampled by the engines, over which the ECs flag the sampled hits. We link $p$=$\beta e$ and $q$=$\gamma e$ linearly to $e$ to collapse the search to a single erasure-bias while preserving relative rates. We add 1Q/\allowbreak 2Q depolarizing channel, $\mathcal{P}(p)$ with probability $p$ and a binary-symmetric flip, $\mathcal{N}(q)$, with probability $q$ over the measurements, $M^{P}$,  where $P \in \{X, Z\}$. These channels are the primitives used to \emph{annotate erasures} based on the EC outcomes as previously explained in Section~\ref{subsec:bg_erasure_qubits}. We then select a schedule $S\in\{S_{\mathrm{4EC}},S_{\mathrm{2EC}}\}$ that fixes the placement of \emph{erasure checks} (ECs). The schedule induces the segment decomposition, introduced in Section~\ref{subsubsec:bg-segments}. For BB codes we compress the seven rounds in Table~\ref{tab:seven_phase_schedule} into $4$ CNOT \emph{bundles}. This 4-bundle partition balances $X$/$Z$ interactions and aligns EC points with depth-1 boundaries so resets don’t create cross-bundle hazards, where a qubit would be reset or measured while it is also scheduled to participate in a CNOT of the next bundle. We place erasure checkpoints $\{A, B, C, D\}$ immediately after each bundle as shown in Figure~\ref{fig:erasure_check_schedule} to create the schedules $S_{\mathrm{4EC}} = S_{\mathrm{ABCD}}$ and $S_{\mathrm{2EC}} = S_{\mathrm{BD}}$. Since checkpoint $D$ coincides with mandatory post-measurement resets, all 2EC schedules include $D$. Among \{$AD,BD,CD$\}, we choose \textbf{$BD$} as it yields balanced segment lengths ($A+B: 3{+}4=7$ phases, $C+D: 3{+}2=5$) and lower worst-case flag latency than $AD$ ($3|9$) or $CD$ ($10|2$). Concretely, $A$=\{$Z_{1},X_{1},Z_{2}$\}, $B$=\{$X_{2},Z_{3},X_{3},Z_{4}$\}, $C$=\{$X_{4},Z_{5},X_{5}$\}, and $D$=\{$Z_{6},X_{6}$\}. Alternatives like checkpoints at $A$ and $C$ are less balanced in both latency and coverage.  Concretely, checkpoint $A$ follows the first bundle, which spans phases \{$Z_{1}, X_{1}, Z_{2}$\}. $B$ follows the second bundle, spanning \{$X_{2}, Z_{3}, X_{3}, Z_{4}$\}. $C$ follows the third bundle, spanning \{$X_{4}, Z_{5}, X_{5}$\} and precedes ancilla readout and $D$ marks the checks at the readout and resets, covering \{$Z_{6}, X_{6}$\}.

For schedule $S$, the compilation stage performs a deterministic lowering of $C_0$ into an erasure-annotated intermediate $C_E$. This ensures both engines consume the same erasure-annotated circuit and enables reproducible head-to-head comparisons. For each segment $s$ of qubit $q$, we enumerate the time-ordered two-qubit interaction events \(G_1,\ldots,G_r\) on the qubit (one per partner in a multi-target bundle) and carry forward the associated space-time location $F_1,\ldots,F_r,F_{r+1}$, as defined in Section~\ref{subsubsec:bg-segments} and Figure~\ref{fig:segments}. Enumerating each two-qubit gate as a distinct event yields canonical fault sites $\{F_i\}$ and cleanly bounds the per-segment site count by $r+1$ for both engines. At each selected checkpoint $\ell\in S$ between consecutive segments $t\!\to\!t{+}1$, an erasure check $\mathcal{E}_{t,\ell}$ emits a flag bit per qubit present at that boundary. The flag is modeled as the output of the classical channel $\mathcal{N}(q)$. We thus define the flagged set $E_{t,\ell}$ and the erasure log paired with the site surfaces $\{F_i\}$ as shown in Eq.~(\ref{eq:erasure_log}):
\begin{equation}
\Lambda=\{(q,t,\ell)\;|\; q\in E_{t,\ell},\ \ell\in S\} \ ,
    \label{eq:erasure_log}
\vspace{-1.5mm}
\end{equation}


%% file: sections/methodology/erasure_approx_exact_injection/erasure_approx_exact_injection.tex
\subsection{Erasure Injection (Exact and Approximate)}\label{subsec:methodology_erasure_approx_exact}

Building on the site surface $\{F_i\}$ defined in the preceding stage, this stage replaces erasure checks/flags with explicit Pauli channels and converts an erasure circuit $C_{E}$ into a decoder-ready stabilizer circuit $C$ using Algorithm~\ref{alg:bibieq-unified}. We convert EC evidence into Pauli noise at canonical sites, remove EC/resets, and output a stabilizer circuit for any decoder, enabling direct Exact vs Approx comparisons \cite{b2,b3}. \emph{\textbf{BiBiEQ-Exact}} preserves the posterior structure implied by erasure checks by sampling a single first hit per qubit/segment and inserting correlated two-qubit channels thereafter, plus a terminal one-qubit channel. \emph{\textbf{BiBiEQ-Approx}} preserves per-location marginals while factorizing correlations into independent single-qubit insertions at partner and terminal sites.

Both engines operate segment-by-segment using the segment abstraction from Section~\ref{subsubsec:bg-segments}. For a segment $s$ on a qubit, let $G_1,\ldots,G_r$ be the entangling gates that act on the qubit inside $s$. The canonical erasure locations are $F_1,\ldots, F_r$ (immediately after each $G_i$ on the partner wire) and the terminal location $F_{r+1}$ on qubit (\textit{lines 1--3}). Hence the only fault-insertion sites for a qubit in $s$ are given as Eq.~(\ref{eq:site_locations_Fi}):
\begin{equation}
    \mathcal{F}(s,q)=\{F_1,\ldots,F_r,F_{r+1}\}, \qquad |\mathcal{F}(s,q)|=r+1.
    \label{eq:site_locations_Fi}
    \vspace{-1.5mm}
\end{equation}

In our seven-phase BB circuits, the schedule bounds $r$: under 4EC when $S_{\mathrm{4EC}} = S_{\mathrm{ABCD}}$, $r\le 2$ so $|\mathcal{F}(s,q)|\le 3$, under 2EC when $S_{\mathrm{2EC}} = S_{\mathrm{BD}}$, $r\le 4$ so $|\mathcal{F}(s,q)|\le 5$. Inputs are the compiled, erasure-annotated seven-phase BB circuit $C_E$, its segment partition $S$ (with implicit 2EC/4EC schedule), the noise law $(e,p,q)$, and the qubit set (data+ancilla).

For each $s\in S$ and qubit, both engines first compute the EC likelihood and the first-hit posteriors $\Pr(A_i\mid d_s)$. Let $d_s\in\{0,1\}$ be the EC outcome (flag=1) (\textit{line 4}). Assuming a homogeneous per-location erasure rate $e$ inside $s$ we calculate the EC outcome using Eq.~(\ref{eq:ec_outcome_prob}):
\begin{equation}
\Pr(d_s{=}1)=\big[1-(1-e)^{r+1}\big](1-q)+(1-e)^{r+1}q \ .
    \label{eq:ec_outcome_prob}
    \vspace{-1.5mm}
\end{equation}
and $\Pr(d_s{=}0)=1-\Pr(d_s{=}1)$. We then define an event $A_i$ as the first erasure event in segment $s$ for qubit between $G_{i-1}$ and $G_i$, where $i=1,\ldots,r{+}1$. The posterior probabilities are calculated using Eq.~(\ref{eq:posterior_prob}):
\begin{equation}
a_i \equiv \Pr(A_i\mid d_s)=
\begin{cases}
\dfrac{e(1-e)^{i-1}(1-q)}{\Pr(d_s{=}1)}, & d_s=1,\\[6pt]
\dfrac{e(1-e)^{i-1}q}{\Pr(d_s{=}0)},      & d_s=0.
\end{cases}
\label{eq:posterior_prob}
\vspace{-1.5mm}
\end{equation}
Equation~\eqref{eq:posterior_prob} gives the first-hit posteriors $a_i$, which is the probability of the first erasure occurring on a qubit between $(G_{i-1},G_i]$ given $d_s$. This turns the EC evidence into per-location probabilities used by both engines to realize Pauli channels at the canonical sites $\mathcal{F}(s,q)$ (\textit{line 5}). With the posteriors $\{a_i\}$ computed, the remainder of the algorithm realizes them via two engines—\emph{\textbf{BiBiEQ-Exact}} and \emph{\textbf{BiBiEQ-Approx}}—described next.

\input{sections/methodology/erasure_approx_exact_injection/bibieq_exact_engine}

\input{sections/methodology/erasure_approx_exact_injection/bibieq_approx_engine}

%% file: sections/methodology/erasure_approx_exact_injection/bibieq_exact_engine.tex
\subsubsection{BiBiEQ-Exact}
\label{subsubsec:bibieq_exact_engine}

\begin{algorithm}[!htbp]
\footnotesize
\caption{BiBiEQ conversion with shared EC$\to$posterior inference and Exact/Approx realization}
\label{alg:bibieq-unified}
\begin{minipage}{\linewidth}
\textbf{Input:} erasure circuit $C_E$, segment partition $S$, EC outcomes $\{d_s\}$, noise $(e,p,q)$, mode $\in\{\textsc{Exact},\textsc{Approx}\}$ \\
\textbf{Output:} stabilizer circuit $C$
\begin{algorithmic}[1]
\State $C \gets C_E$
\For{segment $s\in S$ on wire $q$}
  \State extract $\{G_i\}_{i=1}^{r}$ and $\{F_i\}_{i=1}^{r+1}$ for $s$
  \State $d_s \gets$ EC outcome for $s$ \hfill  \textbf{// (Eq.~\eqref{eq:ec_outcome_prob})}
  \State $\mathbf{a} \gets \textsc{Posteriors}(d_s; e,q)$ \hfill \textbf{// (Eq.~\eqref{eq:posterior_prob})}
  \If{\textsc{Exact}}
    \State $\mathbf{b} \gets \textsc{IndependentProbs}(\mathbf{a})$ \hfill \textbf{// (Eq.~\eqref{eq:bernoulli_var_eq})}
    \State $k \gets \textsc{FirstHit}(\mathbf{b})$  \hfill
    \For{$j=k$ \textbf{to} $r$}
      \State \textsc{Insert2Q}$(C,F_j,\textsc{P\_Exact}(j,\mathbf{b}))$ \hfill \textbf{// (Eq.~\eqref{eq:per_loc_error_prob_exact_engine})}
    \EndFor
    \State \textsc{Insert1Q}$(C,F_{r+1},\textsc{P\_Exact}(r{+}1,\mathbf{b}))$ \hfill \textbf{// (Eq.~\eqref{eq:per_loc_error_prob_exact_engine})}
    \EndIf
  \If{\textsc{Approx}}
    \State $\bar{\mathbf{a}} \gets \textsc{Cumulative}(\mathbf{a})$ \hfill \textbf{// (Eq.~\eqref{eq:posterior_approx_engine})}
    \For{$j=1$ \textbf{to} $r$}
      \State \textsc{Insert1Q\_Partner}$(C,F_j,\textsc{P\_Approx}(\bar a_j))$ \hfill \textbf{// (Eq.~\eqref{eq:approx_engine_probability})}
    \EndFor
    \State \textsc{Insert1Q}$(C,F_{r+1},\textsc{P\_Approx}(\bar a_{r+1}))$ \hfill \textbf{// (Eq.~\eqref{eq:approx_engine_probability})}
  \EndIf
\EndFor
\State \textsc{Drop\_EC\_Reset}$(C)$
\State \Return $C$
\end{algorithmic}
\end{minipage}
\end{algorithm}

Using the posterior probabilities $\{a_i\}_{i=1}^{r+1}$ from Eq.~\eqref{eq:posterior_prob} and keeping the abstractions of the segments intact we compute the independent Bernoulli variables $B_i$ with success probabilities $b_i$ for $i=1,\ldots,r{+}1$ given by  Eq.~(\ref{eq:bernoulli_var_eq}):
\begin{equation}
    b_1=a_1,\qquad
b_i=\frac{a_i}{\prod_{j<i}(1-b_j)}\quad (i\ge 2) \ .
\label{eq:bernoulli_var_eq}
\vspace{-1.5mm}
\end{equation}
By Lemma~1 \cite{b3}, sampling the $B_i$ and taking $k=\min\{i:\,B_i=1\}$ is distributionally equivalent to sampling the disjoint first-hit events $A_i$ with the convention $k=r{+}1$ if no $B_i$ with $i\le r$ fires (\textit{lines 6--7}). This independent-mechanism form turns a global first-hit event into local per-location draws. It preserves the suffix correlations and the single-site marginals, and it allows us to compute channels in a single forward pass with closed-form per-site rates, keeping the procedure linear-time and schedule-agnostic.

Next, we make a single pass over $s$ and act only on the suffix $j\!\ge\!k$ since the first-hit at $k$ affects only the later sites. Skipping $j<k$ avoids double-counting. When $j\le r$, we insert a two-qubit Pauli channel at $F_j$ on the source qubit and its target. At $F_{r+1}$, we insert a one-qubit Pauli channel on the source qubit \cite{b12}. Let $\overline{F}_j:=\{F_j,\ldots,F_{r+1}\}$ denote the suffix of locations starting at $F_j$. The per-location mechanism probability for all $j=1,\ldots,r{+}1$ is shown by Eq.~(\ref{eq:per_loc_error_prob_exact_engine}):
\begin{equation}
    p^{\text{(Exact)}}_j \;=\; \tfrac{1}{2}\;-\;\tfrac{1}{2}\bigl(1-b_j\bigr)^{\,2^{\,1-2\,|\overline{F}_j|}} \ ,
    \label{eq:per_loc_error_prob_exact_engine}
    \vspace{-1.5mm}
\end{equation}
which matches the single-site marginals implied by the first-hit model and preserves its induced correlations across the suffix, (\textit{lines 8--13}). After all segments are processed, erasure checks and resets are removed, yielding the stabilizer circuit $C$. The pass is a single stream per segment. The 2EC/4EC schedule only bounds the number of sites per segment.

%% file: sections/methodology/erasure_approx_exact_injection/bibieq_approx_engine.tex
\subsubsection{BiBiEQ-Approx}
\label{subsubsec:bibieq_approx_engine}

Unlike BiBiEQ-Exact, which preserves the single–first-hit suffix correlations, this engine factorizes across sites. This retains the exact per-site marginals but can bias joint (multi-site) error rates within a segment. In return it enables a linear-time, single-pass, schedule-agnostic realization with higher throughput.

Using the posteriors $\{a_i\}_{i=1}^{r+1}$ from Eq.~\eqref{eq:posterior_prob}, for each site we compute $F_j$ using Eq.~(\ref{eq:posterior_approx_engine}), (\textit{lines 14--15}):
\begin{equation}
    \bar a_j := \Pr(\text{qubit erased before }G_j \mid d_s)=\sum_{i=1}^{j} a_i,
\ j=1..,r{+}1 \ ,
\label{eq:posterior_approx_engine}
\vspace{-1mm}
\end{equation}
where $\bar a_j$ is the cumulative posterior that a qubit has already been erased by the time we reach $G_j$. It is exactly the single-site marginal at $F_j$ that the approximate engine preserves by converting $\bar a_j$ into an independent per-site depolarizing rate. We then approximate the effect of erasures in $s$ by placing, at each $F_j$, a \emph{one-qubit depolarizing channel} whose error probability equals $3\bar a_j/4$. We include three independent error mechanisms at $F_j$. These constitute the nontrivial Pauli operators $P\in\{X,Y,Z\}$ with probability calculated using Eq.~(\ref{eq:approx_engine_probability}), (\textit{lines 16--21}):
\begin{equation}
    p^{\text{(Approx)}}_j \;=\; \tfrac{1}{2}\Bigl(1-\sqrt{\,1-\bar a_j\,}\Bigr) \ ,
    \label{eq:approx_engine_probability}
    \vspace{-1.5mm}
\end{equation}

For $j\le r$, the channel acts on the partner wire at $F_j$ (just after $G_j$) at $F_{r+1}$ it acts on a qubit.

Realization proceeds once per segment in program order with independent per-site draws. This preserves the per-location marginals $\{\bar a_j\}$ implied by the EC evidence but discards their joint dependence across locations. In practice, this can under- or over-estimate multi-site events relative to BiBiEQ-Exact while keeping the same single-site statistics and offering higher throughput. After executing the engines, we drop the EC operations and resets (\textit{line 22}), yielding a standard stabilizer circuit $C$ for downstream decoding and computation of per-round logical error rates.

%% file: sections/performance/Performance.tex
\section{Experimental Evaluation}
\label{sec:performance}

\input{sections/performance/experimental_setup/exp_setup}

\input{sections/performance/simulation_result_analysis/sim_result_analysis}

%% file: sections/performance/experimental_setup/exp_setup.tex
\subsection{\textbf{Experimental Setup}}
\label{sec:perf:setup}

\input{sections/performance/experimental_setup/eval_benchmark}

\input{sections/performance/experimental_setup/eval_metrics}

%% file: sections/performance/experimental_setup/eval_benchmark.tex
\subsubsection{\textbf{Benchmarks for Evaluation}}
\label{subsubsec:bench_eval}

We evaluate the three Bivariate Bicycle (BB) codes from Table~\ref{tab1:bb_codes_family}. For each code we instantiate the seven-phase memory circuit in the $X-$ basis with the number of syndrome-measurement cycles set equal to the code distance $d$. Each circuit is compiled under two erasure-check schedules, 2EC and 4EC \cite{b1,b2,b3}. From the resulting erasure circuit $C_E$, we produce stabilizer circuits $C$ using the two engines: \textit{BiBiEQ-Exact} and \textit{BiBiEQ-Approx}.

For each configuration of code, schedule, and engine, we sweep the erasure probability $e$ over the predefined schedule in the circuit. The error rates Pauli $p$ and the measurement faults $q$ are included and scaled with $e$ as $p=0.1e$ and $q=e$ \cite{b1,b3}. For each configuration we sample $S\!\ge\!2{,}000$ erasure circuits $C_E^{(s)}$. Each $C_{E}^{(s)}$ is converted into the stabilizer circuit $C$ using either of the engines and then simulated using \texttt{Stim} \cite{b11} for at least $500$ shots, yielding atleast $10^6$ shots per configuration, enabling stable $p_L$ estimates \cite{b2, b3}. We increase the sampling count and shot count when choosing a larger BB code from Table~\ref{tab1:bb_codes_family} to achieve $\le10^{-6}$ LER. Decoding uses BP+OSD with the default \texttt{sinter} configuration (sum–product BP, max 30 iterations, OSD of order 60), held constant across all experiments \cite{b13}. For every stabilizer circuit $C$ we record shots, errors, and discards that hold excluded runs due to flagged invalid syndromes.


%% file: sections/performance/experimental_setup/eval_metrics.tex
\subsubsection{\textbf{Evaluation Metrics}}
\label{sec:perf:metrics}

We first compute the \emph{block-level} logical error rate as it denotes the outcome of each full memory experiment. Fixing the code, schedule, engine, and e  we aggregate over the $S$ sampled stabilizer circuits $C^{(s)}$ by summing errors ($E$) and effective shots ($N$): 
\[
E=\sum_{s=1}^{S}\mathrm{errors}^{(s)},\qquad 
N=\sum_{s=1}^{S}\Bigl(\mathrm{shots}^{(s)}-\mathrm{discards}^{(s)}\Bigr),
\vspace{-1.5mm}
\]
We determine the block error rate $p'_L = E/N$. Using $p'_L$ across a block spanning $d$ syndrome-measurement rounds supports fair comparisons across engines and schedules even in the presence of discards. To compare results across code distances and to enable subthreshold-scaling fits \cite{b4}, we subsequently convert the block error rate to a \emph{per-round} logical error rate given as Eq.~(\ref{eq:per_rd_ler}):
\begin{equation}
    p_L \;=\; 1-\bigl(1-p'_L\bigr)^{1/d} \ ,
    \label{eq:per_rd_ler}
    \vspace{-1.5mm}
\end{equation}
which preserves curve ordering across engines/schedules, and expresses performance on a per-round timescale relevant for scaling laws and threshold localization \cite{b2, b3}. While $p_L$ is our primary result, we report three additional metrics--engine gap, gap growth rate and geometric-mean separation--to evaluate the engine agreement, sensitivity to $e$, and improvement with distance over the configurations (code, schedule, engine, $e$). We quantify engine agreement using the engine gap $\rho(e)$ shown in  Eq.~(\ref{eq: enginegap_gap_growth}a), and its $e$–sensitivity using the gap growth rate $s(e)$ shown in Eq.~(\ref{eq: enginegap_gap_growth}b):

\begin{equation}
    \label{eq: enginegap_gap_growth}
    (a) \ \rho(e)\!\equiv\! p_L^{\mathrm{Approx}}(e)/p_L^{\mathrm{Exact}}(e); \ 
    (b) \ s(e) \equiv \frac{\mathrm{d}\log \rho(e)}{\mathrm{d}\log e} \ ,
\end{equation}
Eq.~\ref{eq: enginegap_gap_growth}(a) gives the gap ratio between engines at fixed $e$. Below the pseudo-threshold, both engines are in close agreement when $\rho\!\approx\!1$, at which the BiBiEQ-Approx engine can be used as a reliable proxy for BiBiEQ-Exact for larger sweeps. Eq.~\ref{eq: enginegap_gap_growth}(b) is the log–log derivative (elasticity) of this gap with respect to $e$ and quantifies how fast BiBiEQ–Approx drifts away from BiBiEQ–Exact as $e$ increases, reducing the reliance on BiBiEQ-Approx. To measure the effect of distance within a fixed schedule and engine, we report the geometric-mean ($\mathrm{gmean}$) separation over $e$ using Eq.~(\ref{eq:distance_separation}):
\begin{equation}
    \Gamma_{A\to B}\!\equiv\! \mathrm{gmean}\!\bigl[p_L(A,e)/p_L(B,e)\bigr] \ ,
\label{eq:distance_separation}
\vspace{-1.5mm}
\end{equation}
Here $A\!\to\!B$ denotes a move from distance $A$ to a higher distance $B$ (e.g., $6\!\to\!10$). Values $\Gamma_{A\to B}\!>\!1$ indicate improvement from increasing distance.

The preceding evaluations of $p_L$ compare behavior at a fixed erasure rate $e$ but do not capture how errors decrease with code distance across the region below the pseudo-threshold (subthreshold region). We use the subthreshold scaling law as shown in Eq.~(\ref{eq:subscale}):
\begin{equation}
\widehat{p}_L(e,d)\;=\;A(e)\,\Bigl(\tfrac{e}{\hat e}\Bigr)^{\alpha d} \ ,
\label{eq:subscale}
\vspace{-1.5mm}
\end{equation}
where $d$ is the code distance, $A(e)$ is the distance-independent baseline at erasure rate $e$, and $\alpha>0$ is the subthreshold exponent governing how strongly each unit increase in $d$ reduces error. We model the distance dependence for $e<\hat e$, where $\hat e$ is the per-round pseudo-threshold error rate for the given (schedule, engine) setting. Taking natural logarithms gives an affine function of $d$ whose slope $\alpha\log(e/\hat e)$ is negative below threshold, as shown in Eq.~(\ref{eq:log_subthreshold_scaling}):
\begin{equation}
    \log \widehat{p}_L(e,d)=\log A(e)+\alpha\,d\,\log\!\Bigl(\tfrac{e}{\hat e}\Bigr) \ ,
    \label{eq:log_subthreshold_scaling}
    \vspace{-1.5mm}
\end{equation}
For each fixed $e<\hat e$, we estimate $\alpha(e)$ and $A(e)$ by weighted least squares (WLS) of $\log p_L(e,d)$ versus $d$, using effective shot counts as weights. We also report a pooled exponent $\bar\alpha$ from a single WLS of $\log p_L$ on $d\,\log(e/\hat e)$ over all selected $(e,d)$, with WLS standard errors. This provides a variance-reduced summary of distance sensitivity that enables consistent comparisons across codes, schedules, and engines.

%% file: sections/performance/simulation_result_analysis/sim_result_analysis.tex
\subsection{\textbf{Simulation Results \& Analysis}}
\label{sec:perf:res_analysis}

\subsubsection{\textbf{Decoding Comparison}}
\label{perf:subsubsec:decoding_comp}
\begin{figure}[!b]
  \centering
  \includegraphics[width=.9\columnwidth, keepaspectratio]{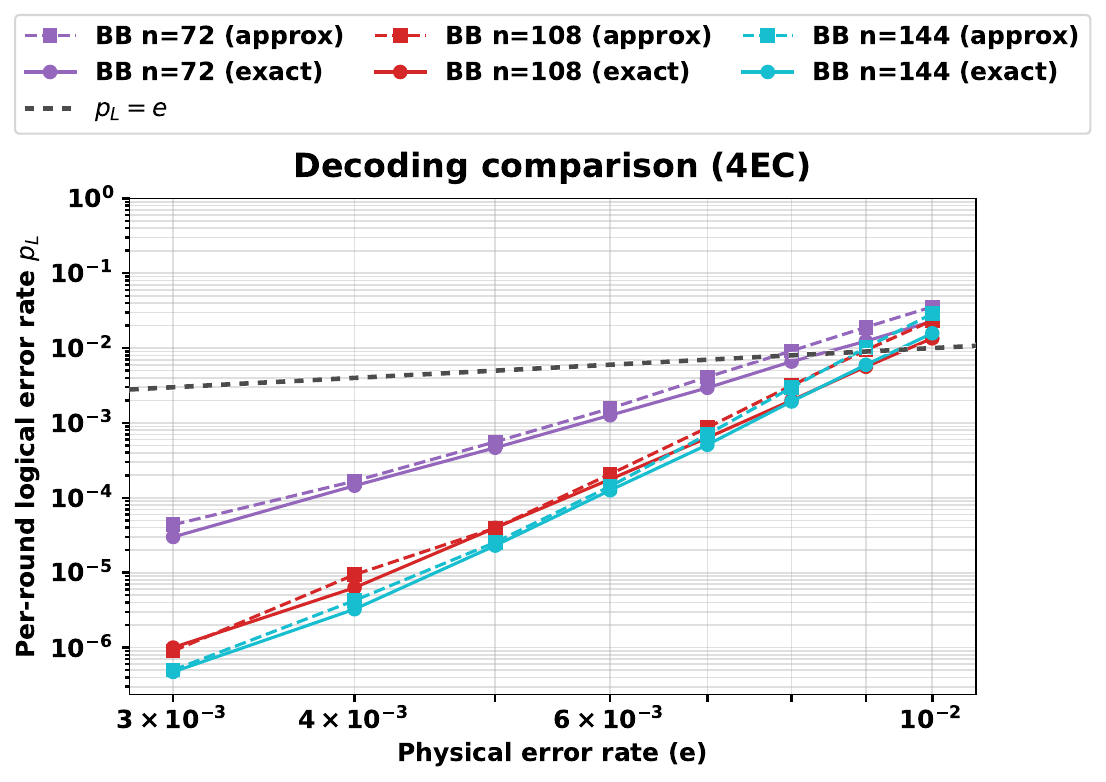}
  \caption{Decoding comparison (4EC): per-round LER $p_L$ vs erasure rate $e$ for BB codes with (solid=Exact, dashed=Approx).}
  \label{fig:bb_ler_4ec}
\end{figure}

\begin{figure}[!t]
  \centering
  \includegraphics[width=.9\columnwidth, keepaspectratio]{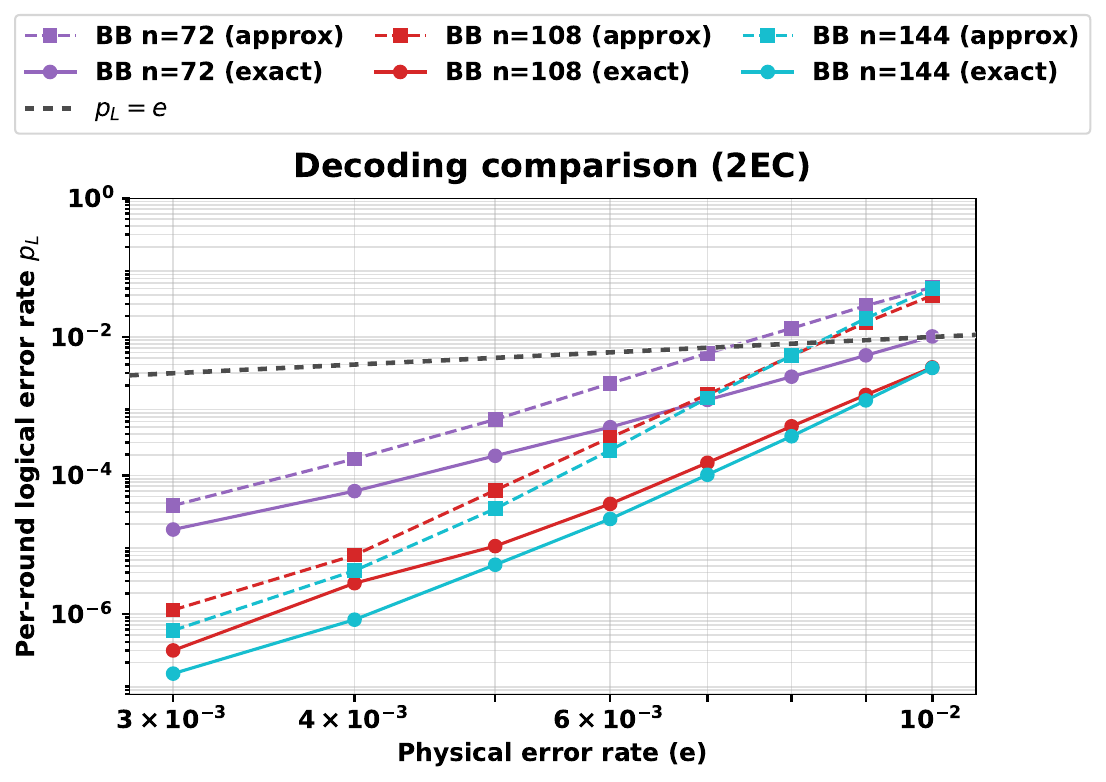}
  \caption{Decoding comparison (2EC): per-round LER $p_L$ vs erasure rate $e$ for BB codes with (solid=Exact, dashed=Approx).}
  \label{fig:bb_ler_2ec}
    \vspace{-4mm}
\end{figure}

We plot the $p_L$ versus $e$ for all BB codes in Table~\ref{tab1:bb_codes_family}, where Figure~\ref{fig:bb_ler_4ec} plots $p_L$ for the 4EC schedule and Figure~\ref{fig:bb_ler_2ec} for the 2EC schedule. For each (schedule, engine), the pseudo-threshold $\hat e$ satisfies $p_L(\hat e)=\hat e$ and defines our operating regime. In both schedules and for both engines, larger BB codes achieve lower per-round $p_L$ at fixed $e$. The curves monotonically increase in $p_L$ across the entire sweep. The separation is most pronounced well below the pseudo-threshold and narrows as $e\!\to\!\hat e$, consistent with the subthreshold scaling fits further explained in Section~\ref{perf:subsubsec:subthreshold_scaling}. The comparisons of scaling and distance are made for $e<\hat e$. In Figure~\ref{fig:bb_ler_4ec}, the \emph{engine gap} calculated from Eq.~(\ref{eq: enginegap_gap_growth}a) shows $\rho(e) \!\approx\!1$ across the sweep, indicating that 4EC suppresses segment correlations. In Figure~\ref{fig:bb_ler_2ec}, the engines separate at higher $e$, with BiBiEQ–Approx drifting upward relative to BiBiEQ–Exact. Under 4EC, the gap is about $3$–$7\times$ smaller than 2EC. The median engine gap across codes falls from roughly $4$–$8$ (2EC) to about $1.2$–$1.3$ (4EC) showing a $70-83\%$ reduction. Using the \emph{gap growth rate} $s$ from Eq.~(\ref{eq: enginegap_gap_growth}b), we analyze how fast BiBiEQ-Approx drifts from BiBiEQ-Exact as $e$ increases. For BiBiEQ-Approx, the gap growth rate drops from $0.85$–$1.36$ (2EC) to $0.02$–$0.39$ (4EC) showing a $\sim 39\times$ slower growth across the operating region.

Across both figures, the exact curve lies below the approx curve, indicating that BP+OSD benefits from the joint-erasure structure preserved by BiBiEQ–Exact. Using the engine gap $\rho$ (Approx/Exact), we observe that the median advantage of BiBiEQ–Exact over BiBiEQ–Approx reaches up to $88\%$ on 2EC and up to $24\%$ on 4EC schedule. BiBiEQ–Approx can be chosen for faster design sweeps with a 4EC schedule but not for a 2EC schedule thus realizing the choice of scheduling to be consequential for BiBiEQ-Approx but marginal for BiBiEQ-Exact as it sets accurate LERs for both schedules.

To isolate the effects of distance at low erasure rate $e$, we use the \emph{distance separation} $\Gamma_{A\to B}$ from Eq.~\eqref{eq:distance_separation}. For $e\le 4\times 10^{-3}$ we find $\Gamma_{6\to10}$ reduces the per-round logical error by a factor of roughly $26$–$34\times$ and $\Gamma_{10\to12}\approx 2.0$–$2.7\times$ reduction, indicating that the $6\!\to\!10$ step delivers most of the gain. This ordering matches the subthreshold scaling law where the reduction factor depends exponentially on the increment $\Delta d$, i.e., $(e/\hat e)^{\alpha\,\Delta d}$, so the larger jump ($\Delta d{=}4$ for $6\!\to\!10$) suppresses errors far more than the smaller jump ($\Delta d{=}2$ for $10\!\to\!12$). In our low-$e$ range, most distance-driven benefit is captured by $d:6\!\to\!10$.

\input{sections/performance/simulation_result_analysis/subthreshold_scaling}

%% file: sections/performance/simulation_result_analysis/subthreshold_scaling.tex
\subsubsection{\textbf{Subthreshold Scaling}}
\label{perf:subsubsec:subthreshold_scaling}

\begin{figure*}[!t]
  \centering
  \begin{subfigure}[t]{.245\textwidth}
    \centering
    \includegraphics[width=\linewidth]{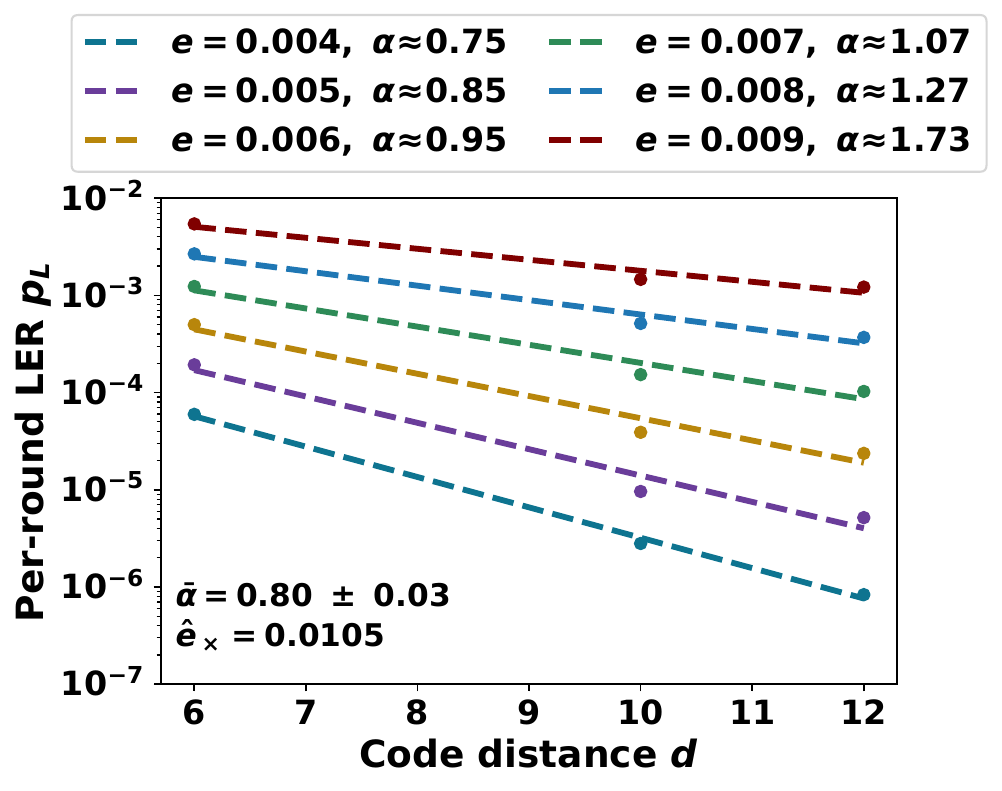}
    \subcaption{2EC : BiBiEQ-Exact}
  \end{subfigure}\hfill
  \begin{subfigure}[t]{.245\textwidth}
    \centering
    \includegraphics[width=\linewidth]{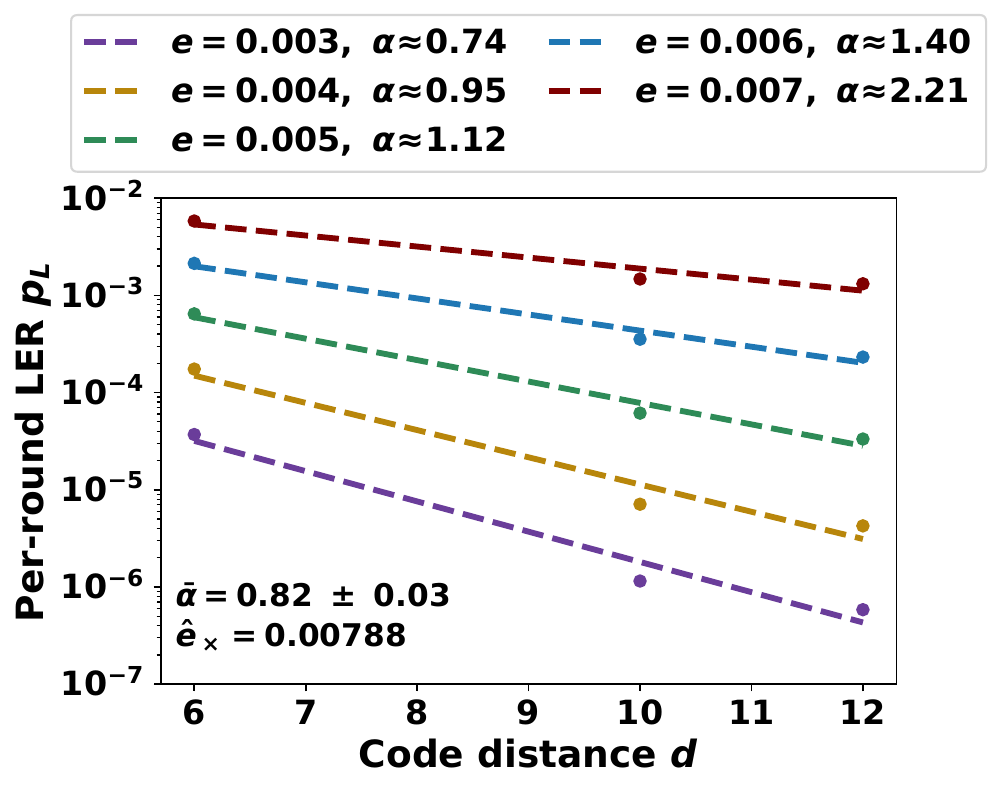}
    \subcaption{2EC : BiBiEQ-Approx}
  \end{subfigure}\hfill
  \begin{subfigure}[t]{.245\textwidth}
    \centering
    \includegraphics[width=\linewidth]{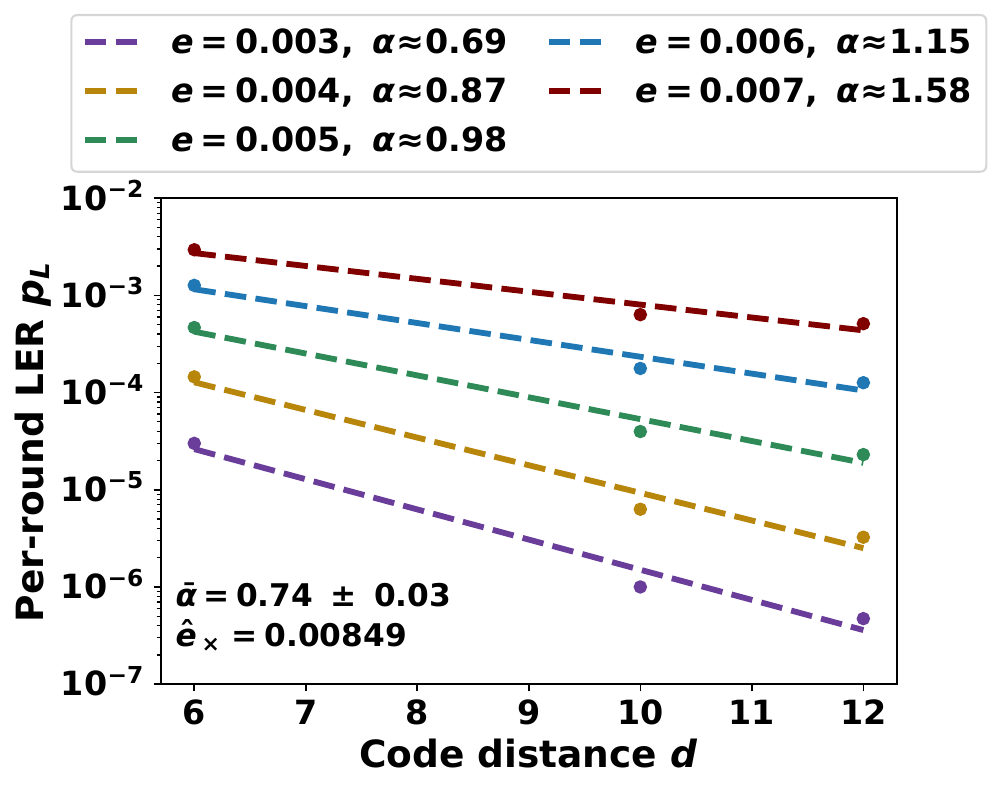}
    \subcaption{4EC : BiBiEQ-Exact}
  \end{subfigure}\hfill
  \begin{subfigure}[t]{.245\textwidth}
    \centering
    \includegraphics[width=\linewidth]{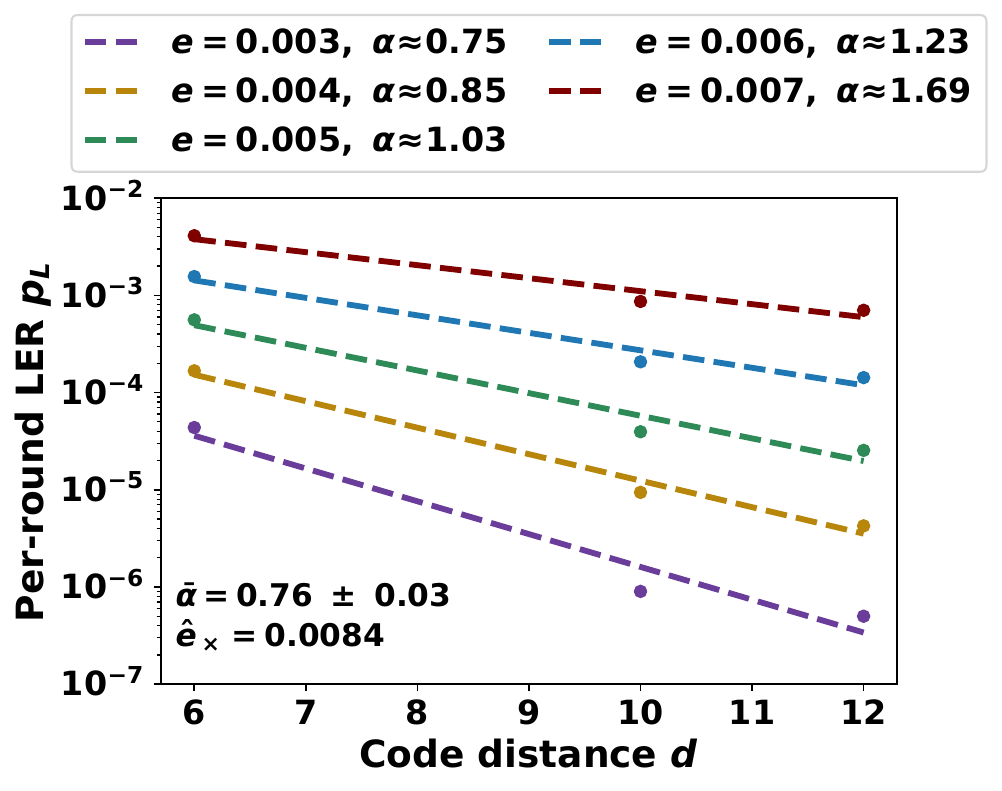}
    \subcaption{4EC : BiBiEQ-Approx}
  \end{subfigure}

\caption{Subthreshold scaling vs.\ code distance $d$ for BB codes under 2EC and 4EC (Exact and Approx).
Markers show measured per-round $p_L$ at $d\in\{6,10,12\}$ for several fixed $e<\hat e$ (one color per $e$); dashed lines are weighted least-squares fits of $p_L \approx A(e)\,(e/\hat e)^{\alpha d}$.
Legends list $e$ and the fitted $\alpha(e)$; panel annotations report the pooled exponent $\bar\alpha$ and the pseudo-threshold $\hat e$ used for the fit.}

  \label{fig:bb_subscale_oneblock}
    \vspace{-6mm}
\end{figure*}

Figure~\ref{fig:bb_subscale_oneblock} instantiates the scaling law of Eqs.~\eqref{eq:subscale}–\eqref{eq:log_subthreshold_scaling} below the pseudo-threshold $\hat e$ (subthreshold region) for each pair of schedule and engine. Panels are arranged by schedule (2EC, 4EC) and engine (BiBiEQ–Exact, BiBiEQ–Approx). Within each panel, markers show measured $p_L$ at $d\in\{6,10,12\}$ for fixed $e<\hat e$ (colors denote $e\in[0.003,0.009]$). The dashed lines represent the WLS fits using the estimation procedure defined in Section~\ref{sec:perf:metrics}. Legends list the fixed $e$ values and the fitted $\alpha(e)$. The panel annotations report the pooled exponent $\bar\alpha$ and the $\hat e$ used in the fit. The near-linear decay in $\log p_L$ versus $d$ reflects the exponential distance suppression predicted by the model, with slope proportional to $\log(e/\hat e)$ for each fixed $e<\hat e$.

Applying the fit to the LER data yields \textbf{two} clear trends:

\emph{\textbf{\underline{Engine effect}:}} The distance exponent is approximately engine-invariant. Under 2EC we obtain \(\bar\alpha\approx 0.80\) (BiBiEQ-Exact) and \(0.82\) (BiBiEQ-Approx) and for 4EC we obtain \(\bar\alpha\approx 0.74\) (BiBiEQ-Exact) and \(0.76\) (BiBiEQ-Approx). Thus, engine choice primarily shifts the distance-independent level $A(e)$ rather than the slope in $d$. The key takeaway is that, similar to BiBiEQ-Exact,  BiBiEQ-Approx preserves the way performance scales with distance. It is therefore reliable for fast design sweeps that compare codes or distances.

\emph{\textbf{\underline{Schedule effect}:}} The schedule adjusts the reference error rate with only mild changes in slope. In the BiBiEQ-Exact pipeline \(\hat e\) moves from \(1.05\times 10^{-2}\) (2EC) to \(8.5\times 10^{-3}\) (4EC), a decrease of about \(19\%\). For BiBiEQ-Approx engine it moves from \(7.9\times 10^{-3}\) (2EC) to \(8.4\times 10^{-3}\) (4EC), an increase of about \(7\%\). The key takeaway is to choose the schedule that maximizes the margin \((\hat e - e)\) for the engine in use. Since 4EC also tightens the engine agreement seen in figure~\ref{fig:bb_ler_4ec}, it is the safer choice when using BiBiEQ-Approx during architecture exploration, while BiBiEQ-Exact can be used to determine the accurate LER for both schedules.

In summary, these findings support the use of BiBiEQ-Approx with 4EC for dependable relative comparisons across codes and distances, while BiBiEQ-Exact remains the benchmark for accurate LER. Within the erasure range studied, the dominant distance-driven improvement arises from the step \(d{:}6\!\to\!10\), with diminishing returns beyond \(d{=}10\).

%% file: sections/conclusion/Conclusion.tex
\section{Conclusion}
\label{conclusion}
In this work, we introduced BiBiEQ, an erasure-aware compilation and conversion framework for BB codes that (i) generates seven-phase memory circuits under erasure-biased noise and (ii) converts those erasure circuits into stabilizer circuits using two engines—BiBiEQ-Exact and BiBiEQ-Approx—offering an accuracy-throughput trade-off. Evaluated across distances $d\in\{6,10,12\}$ and 2EC/4EC schedules, BiBiEQ consistently reduces the per-round logical error rate with distance. At low erasure $e\le 4\times10^{-3}$, the hop $d{:}6\!\to\!10$ cuts $p_L$ by \textbf{26–34$\times$} and $d{:}10\!\to\!12$ by a further \textbf{2.0–2.7$\times$}. Under 4EC, the Approx/Exact engine gap \textbf{shrinks by 70–83\%} relative to 2EC and its gap growth rate with $e$ slows by \textbf{$\sim$39$\times$}. BiBiEQ-Exact outperforms BiBiEQ-Approx by up to \textbf{88\% (2EC)} and \textbf{24\% (4EC)}. Pseudo-thresholds align with these trends, at $\hat e\!\approx\!1.05\times10^{-2}$ (2EC, Exact) vs.\ $8.5\times10^{-3}$ (4EC, Exact) and $7.9\times10^{-3}$ (2EC, Approx) vs.\ $8.4\times10^{-3}$ (4EC, Approx). These results establish BiBiEQ as a practical pathway to erasure-aware QLDPC analysis and decoding in the pseudo-threshold regime, clarifying how schedule choice and engine, shape accuracy and cost. Future research includes integrating erasure-native decoders (\textit{e.g.}, peeling/erasure-matching, union-find), extending the framework to morphing circuits, and applying BiBiEQ across broader QLDPC families.